# Evidence of strong relationship between hemispheric asymmetry in solar coronal rotation and solar activity during solar cycle 24

Jaidev Sharma[1], Anil K Malik[1], Brajesh Kumar[2] and Hari Om Vats[3]

[1]Department of Physics, Chaudhary Charan Singh University, Meerut, 250004, India.
[2]Udaipur Solar Observatory, Physical Research Laboratory, Dewali, Badi Road, Udaipur, 313004, India.
[3]Space Education and Research Foundation, Ahmedabad, 380054.

**ABSTRACT**

In this article, we report an evidence of very high and statistically significant relationship between hemispheric asymmetry in solar coronal rotation rate and solar activity. Our approach is based on cross correlation of hemispheric asymmetry index (AI) in rotation rate with annual solar activity indicators. To obtain hemispheric asymmetry in solar rotation rate, we use solar full disc (SFD) images at 30.4 nm, 19.5 nm and 28.4 nm wavelengths for 24th Solar Cycle i.e., for the period from 2008 to 2018, as recorded by the Solar Terrestrial Relations Observatory (STEREO) space mission. Our analysis shows that hemispheric asymmetry in rotation rate is high during the solar maxima from 2011 to 2014. On the other hand, hemispheric asymmetry drops gradually on both sides (i.e., from 2008 to 2011 and from 2014 to 2018). The results show that asymmetry index (AI) leads sunspot numbers by ~ 1.56 years. This gives a clear indication that hemispheric asymmetry triggers the formation of sunspots working together with the differential rotation of the Sun.

## 1 INTRODUCTION

Solar activity affects the terrestrial climate and space weather (Hathaway, Wilson & Reichmann 1999; Hathaway & Wilson 2004; Rozelot 2001; Hiremath & Mandi 2004; Georgieva et al. 2005). Solar activity indicators like H-alpha flare index, X-ray flares, solar active prominences (SAP), solar proton events, monthly mean sunspot area and monthly mean sunspot numbers on both the hemispheres show asymmetric behavior. The filaments appearance on both the hemispheres during 1964 to 1974 depicted North-South asymmetry (Hansen & Hansen, 1975). Signatures of North-South asymmetry in various solar phenomena were reported by Verma (1993) from 8–22 solar cycles. Sokoloff and Nesme-Ribes (1994) observed high values in North-South asymmetry during the Maunder minimum.
X-ray flare index and solar active prominences (SAP) during 21, 22 and 23 solar cycles showed that southern hemisphere dominates during these solar cycles (Joshi & Joshi 2004; Verma 2000).

Sun's differential rotations produce stretches and twists in the solar magnetic fields, as a result create sunspots and most of the solar activities on its surface. It is well known that complex solar dynamo process controls the solar activities (Cameron, Dikpati & Brandenburg 2017). Thus, the study of evolution of solar rotation with the progress of solar activity cycle is especially important in solar astronomy. Li *et al.* (2014) demonstrated that solar activity and equatorial rotation rate lies in opposite phase. Antiphase correlation between equatorial rotation rate of solar corona using solar x-ray images from 1992 to 2001(except for the year 2000) and SSNs has been reported by Chandra, Vats & Iyer (2010). A study by Jurdana- Sepic *et al.* (2011) on the rotation of the small bright coronal structures (SBCS) on monthly and yearly temporal scales and its relationship with solar activity showed that the solar rotation velocity at equator is lower around solar maxima in the 23rd cycle. Some models by Brun (2004); Brun, Miesch, & Toomre (2004) and Lanza (2007) predicted that the Sun has less differential rotational gradient at solar maxima than that of at solar minima. For solar radio flux at 2.8 GHz, almost random trend of rotation period with respect to sunspot cycles during 1947–2009 is reported (Chandra & Vats; 2011). Recently, Lekshmi, Nandy and Antia (2018) analyzed 16 years of photospheric Doppler images covering the time period from 2001 to 2017 obtained by Global Oscillation Network Group (GONG) and showed that hemispheric asymmetry corresponding to near-surface torsional oscillation velocity is correlated with the respective asymmetry in the magnetic flux as well as sunspot numbers at the surface of the Sun. They also concluded that hemispheric asymmetry in these torsional oscillations leads the asymmetries with respect to photospheric magnetic flux and sunspot numbers.

Here, we report a unique approach to obtain a strong, statistically significant relationship between hemispheric asymmetry in solar coronal rotation rate and solar activity indicators. The results have extremely low standard errors. We use solar full disc (SFD) STEREO-A images at 30.4 nm, 19.5 nm and 28.4 nm wavelengths for 24th solar cycle i.e. period from 2008 – 2018. From daily SFD images at different wavelengths, we extract the mean North and South hemispheric fluxes. Thus, a time series are made for North and South hemispheres corresponding to each year, the flux modulation of these series gives the information about rotation of corresponding hemispheres. Each hemispheric time series of EUV flux is auto correlated up to a shift of 150 days. The first secondary peak of auto correlogram is fitted by Gaussian function (Sharma *et al.* 2020,). The shift corresponding to the peak value of Gaussian function fitted on first secondary maxima is the hemispheric synodic rotation period.

**e-mail:** jaidevsharma24@gmail.com

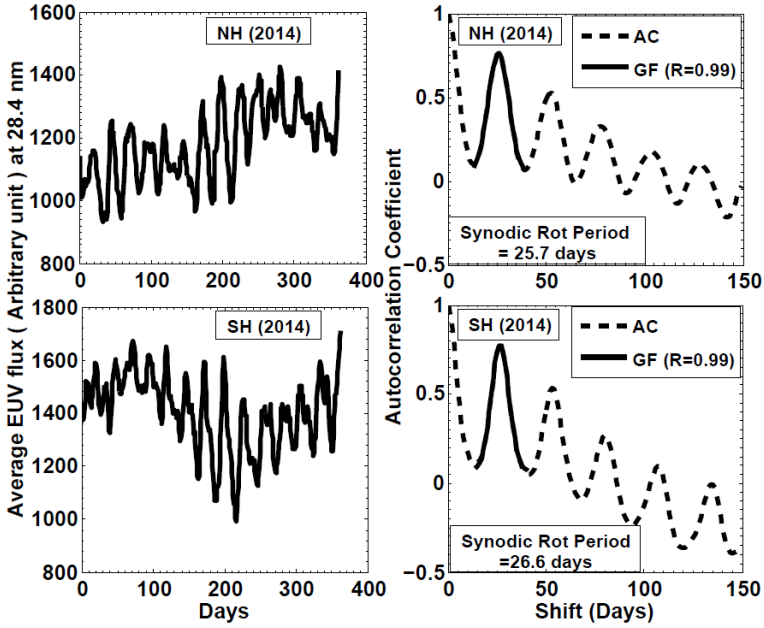

**Figure 1**: The left panels shows variation in averaged EUV flux (for the year 2014) extracted from northern and southern hemispheres using observations at 28.4 nm from *STEREO* while the right panels show corresponding autocorrelograms (Dashed line) with Gaussian fitting (Continuous line) on first secondary maxima.

## 2 DATA REDUCTION AND METHODOLOGY

The *STEREO* space mission is a solar mission of National Aeronautics and Space Administration (NASA), which is observing the Sun since October 2006. Two spacecrafts *(STEREO-A and STEREO-B)* are orbiting around the Sun and obtaining the stereoscopic images of the solar coronal layers as solar full disc (SFD) in the range of multiple wavelengths, viz 30.4 nm, 17.1 nm, 19.5 nm and 28.4 nm. It is also observing solar activity phenomena like coronal mass ejections (CMEs), solar wind and solar plasma. Connection with STEREO-B has been lost in 2014 while STEREO-A is still observing the Sun. The detailed information of this space mission is available at the websites http://stereo.gsfc.nasa.gov/ and http://stereo.jhuapl.edu/.

In this work, we started with the observations at 30.4 nm, 19.5 nm and 28.4 nm wavelength, in order to estimate the hemispheric asymmetry index in solar coronal rotation rate. There is a gap of observations of almost first six months in all the wavelengths in the year 2015 because at that time the STEREO-A spacecraft was passing behind the Sun from the point of view of the Earth, so, ground stations were not able to receive radio signals from the spacecraft and, in fact, most of the instruments were turned off. Hence, there is a gap of STEREO EUV data in our analysis during the aforementioned time period. We extract mean hemispheric EUV flux from SFD images taken one per day by rigorous programming in Interactive data language (IDL). We formed two time series (separately for northern and southern hemispheres) for each year (from 2008 – 2018).These time series of extracted EUV flux have the information of rotational features of the corresponding hemispheres. Sharma *et al.* (2020), Vats *et al.* (2001) and Vats & Chandra (2011) used autocorrelation analysis as one of the established tools to obtain periodicity in any time series. We auto correlate each hemispheric time series of EUV flux up to a shift of 150 days. A typical diagram for averaged hemispheric EUV flux (Left panels) and their corresponding autocorrelogram for the year 2014 at 28.4 observation nm are shown in Fig.1. The right panels in Fig.1 represents the autocorrelograms up to shifts of 150 days for both the hemispheres for the year 2014 using 28.4 nm observations from STEREO /EUV instrument. From Fig.1, it is evident that the first peak in each autocorrelogram has higher value of correlation coefficient as compared to rest of the peaks that clearly shows strong periodicity in rotational features. To estimate accurate rotation period, we fit Gaussian function (Continuous line) to the first secondary peak because first peak is smoother and higher than other peaks. The shift of the peak of the fitted Gaussian function of first secondary maxima is the hemispheric synodic rotation period. The obtained Pearson coefficient R is 0.99, which shows that the fitting is statistically very accurate for the measurements of rotation period. Standard errors in fitting of Gaussian function are extremely small (Table 1), which mean high accuracy in the estimation of rotation periods. In Table 1, two values (in bold) for southern hemisphere are obtained by linear interpolation. Their autocorrelograms had no rotational peak as features in the corresponding images are missing. The hemispheric synodic rotation period is converted into hemispheric sidereal rotation period by the following relation.

$$T_{sidereal} = \frac{346\, T_{synodic}}{346 + T_{synodic}} \quad (1)$$

Here, the orbital period of STEREO-A is 346 days.

The relation between sidereal rotation period (in days) and sidereal rotation rate (in deg/day) is calculated using

$$T_{sidereal} = \frac{360°}{\omega_{rot}} \quad (2)$$

Hence $\omega_{rot}$ (sidereal rotation rate) in deg/day is obtained for both the hemispheres. In this analysis, we use three EUV wavelengths (30.4 nm, 19.5 nm and 28.4 nm) having their corresponding temperatures on logarithmic scale as $\log_{10}$ (4.9), $\log_{10}$ (6.15) and $\log_{10}$ (6.3) Kelvin respectively (Wuelser *et al.* 2004).

We incorporate a new and unique parameter named as Asymmetry Index (AI) to mathematically measure the North-South hemispheric asymmetry in rotation rate, which is defined as

$$AI = \frac{\omega_N - \omega_S}{\omega_N + \omega_S} \quad (3)$$

Where $\omega_N$ represents the rotation rate of northern hemisphere (NH) and $\omega_S$ as that of southern hemisphere (SH) in deg/day. We cross correlate AI with sunspot numbers and EUV flux. To ascertain statistical significance of the results, we applied t-test.

Table 1: The hemispheric synodic rotation periods (in days) with corresponding standard errors (in parentheses) for *STEREO* observations at 30.4 nm, 19.5 nm and 28.4 nm from 2008 to 2018.

| Years | 30.4 nm NH | 30.4 nm SH | 19.5 nm NH | 19.5 nm SH | 28.4 nm NH | 28.4 nm SH |
|---|---|---|---|---|---|---|
| 2008 | 27.14 (0.09) | 27.57 (0.02) | 27.47 (0.02) | 27.37 (0.008) | 26.92 (0.05) | 27.21 (0.001) |
| 2009 | 26.44 (0.084) | **28.02 (0.04)** | 27.08 (0.013) | 27.89 (0.026) | 26.40 (0.012) | 27.19 (0.027) |
| 2010 | 28.10 (0.006) | 28.47 (0.054) | 27.82 (0.002) | 27.96 (0.072) | 27.70 (0.009) | 27.69 (0.015) |
| 2011 | 26.15 (0.064) | 28.97 (0.019) | 25.85 (0.082) | 28.18 (0.012) | 26.35 (0.052) | 27.76 (0.009) |
| 2012 | 25.93 (0.013) | 27.44 (0.035) | 25.94 (0.005) | 27.30 (0.04) | 26.29 (0.035) | 27.59 (0.026) |
| 2013 | 27.17 (0.048) | 28.46 (0.006) | 26.76 (0.025) | 28.48 (0.155) | 26.80 (0.024) | 27.30 (0.009) |
| 2014 | 25.28 (0.018) | 26.24 (0.033) | 25.04 (0.002) | 26.16 (0.034) | 25.69 (0.009) | 26.58 (0.001) |
| 2015 | 24.61 (0.017) | 24.06 (0.024) | 27.89 (0.107) | **26.62 (0.05)** | 27.08 (0.03) | 27.35 (0.021) |
| 2016 | 26.79 (0.025) | 25.61 (0.09) | 27.02 (0.019) | 27.08 (0.061) | 26.74 (0.019) | 25.14 (0.11) |
| 2017 | 27.22 (0.01) | 26.29 (0.069) | 27.29 (0.02) | 26.80 (0.014) | 27.23 (0.019) | 26.87 (0.003) |
| 2018 | 27.53 (0.098) | 26.55 (0.132) | 26.74 (0.062) | 27.89 (0.007) | 26.48 (0.09) | 27.47 (0.045) |

## 3 RESULTS AND DISCUSSIONS

We used SFD images at 30.4 nm, 19.5 nm and 28.4 nm wavelengths, having their respective temperatures as $\log_{10}(4.9)$, $\log_{10}(6.15)$ and $\log_{10}(6.3)$ Kelvin to investigate the temporal variation of hemispheric asymmetry of solar coronal rotation (Fig. 2). The emission at 30.4 nm wavelength belongs to chromosphere-coronal transition region whereas at 19.5 and 28.4 nm wavelengths are from solar corona. The following unique results are obtained.

### 3.1 ASYMMETRIC BEHAVIOUR OF SOLAR CORONA

Hemispheric rotational asymmetry corresponding to 30.4 nm, 19.5 nm and 28.4 nm is highest in 2011 (beginning of the maximum phase of solar activity for $24^{th}$ solar cycle). The AI indicates that during $24^{th}$ solar cycle, the northern hemisphere rotates faster than southern hemisphere. The values of AI for 30.4 nm and 19.5 nm are 4.7% and 4% respectively, whereas that for 28.4 nm it is slightly low (~ 2.4%). The average asymmetry from all three wavelengths (19.5 nm, 30.4 nm and 28.4 nm) follows almost similar temporal variation as that of individual layers. Highest value (3.7%) of average asymmetry indicate higher rotation rate in the northern hemisphere in the year 2011. AI for other years is lower, which indicate a reasonable similarity with the variation of solar activity. This investigation show overall northern hemisphere has higher asymmetry than southern hemisphere. The negative values of AI indicate that rotation rate in southern hemisphere is higher than that in the northern hemisphere. In 2016 (solar minimum phase), AI values are about -2.8 % and -2.1%, for emissions at 30.4 nm and 28.4 nm, respectively. For the emissions at 19.5 nm, AI is least (about 0.9%) in the year 2015. During the ascending phase of the $24^{th}$ solar cycle (i.e. from 2008 to 2010), average AI values are 1.3%, 0.4 % and 0.6 % corresponding to 30.4 nm, 19.5 nm and 28.4 nm wavelengths, respectively.

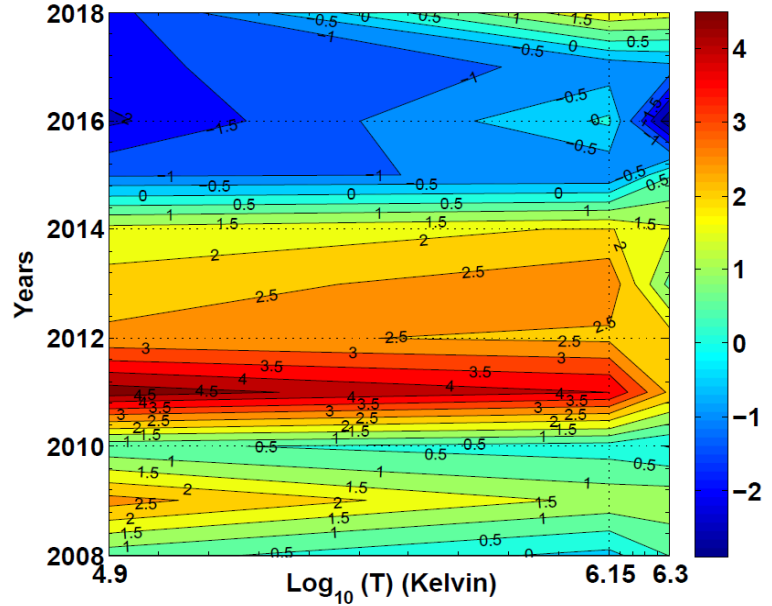

**Figure 2**: The contour plot of the temporal variation of hemispheric asymmetry index (AI) as a function of temperature using observations of the STEREO space mission. It is seen that the AI is high during the solar maximum (2010-2013) phase of the solar cycle 24. The bar on right shows the color code for AI (%).

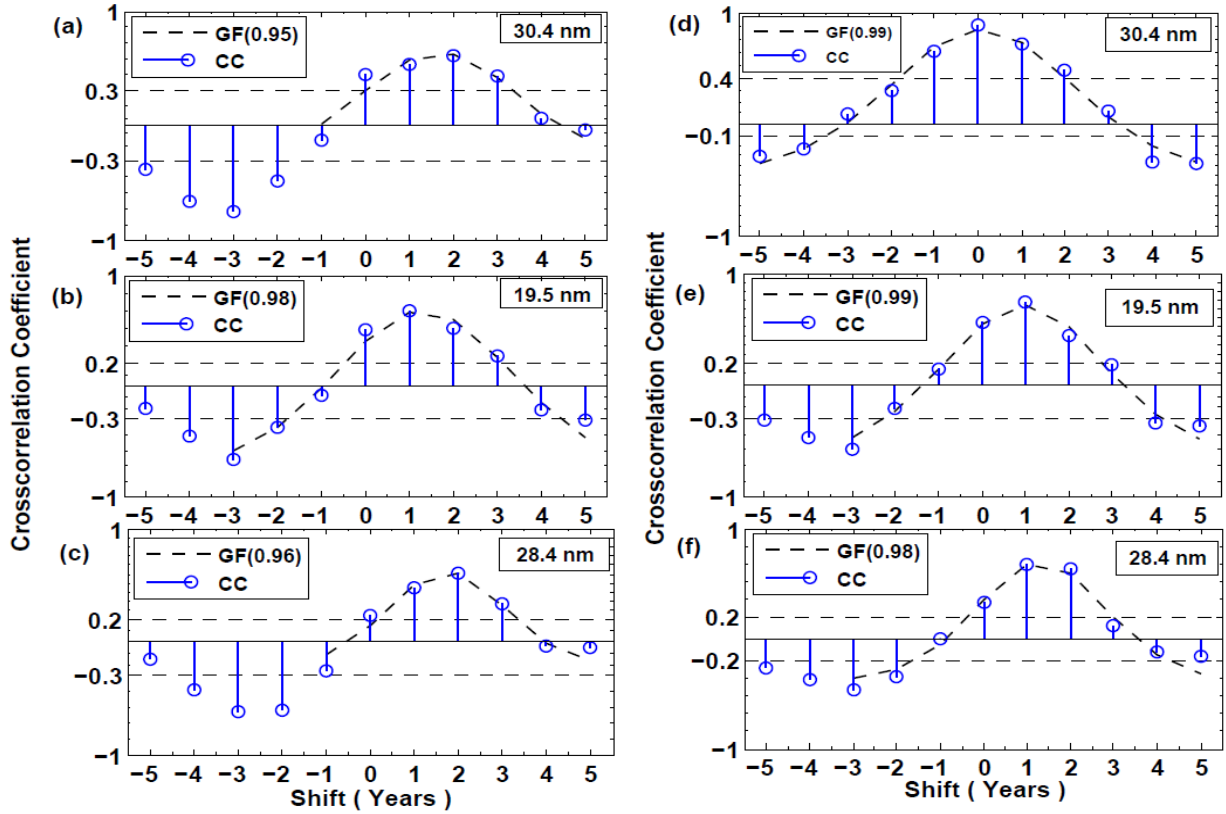

**Figure 3:** Cross-correlation of hemispheric asymmetry index (AI) in rotation rate for 30.4 nm, 19.5 nm and 28.4 nm and annual sunspot numbers as a function of shift (years) are shown in the panels (a),(b) and (c). The panels (d), (e) and (f) depict the cross-correlation of AI in rotation rate for 30.4 nm, 19.5 nm and 28.4 nm and the individual extracted EUV flux as a function of shift in years. The horizontal dashed lines show 95% confidence levels.

Average values of AI increase up to 2.8%, 2.8% and 1.8% during the maximum phase of solar activity cycle from 2011 to 2014. During decaying phase of the cycle from 2015 to 2018, the average values of AI again drop to 1.6%, 0.06% and 0.03%, respectively. Averages AI of all three wavelengths have their values about 0.8%, 2.4% and 0.6% during ascending, maximum and descending phases respectively. This systematic trend shows that the asymmetry follows the solar activity during cycle 24. However, long term study of similar data for more solar cycles is recommended to further ascertain this systematic trend. We use cross-correlation as statistical tool to measures the similarity of AI series at all three wavelengths. We find a strong correlation (about 70%) in them. Thus, asymmetric behavior of transition layer (30.4 nm) and both coronal layers (19.5 nm and 28.4 nm) is similar. The cross correlation of AI (using 30.4nm and 19.5 nm) is even higher (about 75 %).

### 3.2 RELATION BETWEEN AI AND SOLAR ACTIVITY CYCLE

To investigate the relationship between hemispheric asymmetry in solar rotation and solar activity, we use a unique approach based on new parameter named as hemispheric asymmetry index (AI) in rotation rate. AI plays a very crucial role in the investigations of a systematic relationship with the solar cycle activity (e.g. variation of sunspot numbers). We calculated cross-correlation of hemispheric asymmetry index of rotation rate with annual sunspot numbers (SSNs) and extracted EUV flux as a function of shift in years. A systematic and strong correlation of AI with SSN and EUV flux is evident (Fig.**3**). The horizontal dashed lines in all panels in Fig.3 show 95% confidence level. T-test is applied for the estimation of significance of correlation. The left panels (a), (b) and (c) in Fig. **3** show the cross-correlation of asymmetry index (AI) in rotation rate with annual sunspot numbers as a function of shift (in years) corresponding to 30.4, 19.5 and 28.4 nm wavelengths, respectively. The respective statistical significance of cross-correlations in these cases is 95% (Table **2**) that shows a very strong correlation between asymmetry index in rotation rate and annual sunspot numbers. The Gaussian function is fitted to the cross-correlation coefficient (Stem plot) to obtain more accurate value of the shift in years. The Gaussian function peaks are obtained at 1.7, 1.3 and 1.7 years for 30.4 nm, 19.5 nm and 28.4 nm respectively. This clearly illustrates that during 24[th] solar cycle from 2008 to 2018, hemispheric asymmetry in rotation rate leads SSNs with shifts of ~1.56 years (average shifts of 1.7 year, 1.3 year and 1.7 year, for EUV wavelengths, respectively).

We found that cross-correlation of asymmetry index (AI) with SSNs and EUV flux shown in all panels in Fig. **3** show very similar behavior. These results are of statistically high significance (95%) (Table **2**). This confirms an extraordinarily strong correlation. The Gaussian fits for panels (d, e and f) in Fig.3 are found at 0.09 year, 0.9 year and 1.3 year for 30.4 nm, 19.5 nm and 28.4 nm wavelengths, respectively.

Table 2: Cross-correlation Coefficient (CC) between (1) asymmetry index (AI) in rotation rate and annual sunspot numbers (SSNs) and (2) asymmetry index (AI) and average extreme ultra-violet flux (EUV) as a function of lag or shift (in years). The results have very good statistical significance.

| Observations | Peak values of CC b/w AI and SSNs | t-calculated | Statistical significance of cross-correlation | Peak values of CC b/w AI and EUV | t-calculated | Statistical significance of cross-correlation |
|---|---|---|---|---|---|---|
| 30.4 nm | 0.61 | 2.31 | 95 % | 0.88 | 5.56 | 99.9 % |
| 9.5 nm | 0.68 | 2.78 | 95 % | 0.74 | 3.30 | 99 % |
| 28.4 nm | 0.61 | 2.31 | 95 % | 0.67 | 2.71 | 95 % |

This signifies that hemispheric asymmetry in rotation rate leads EUV flux with a shift of about one year. In all cases, the Pearson's coefficient (R) of Gaussian fitting (GF) is more than 0.94 that reflects high accuracy in the shift measurement.

## 4 CONCLUSIONS

Our findings support the results of Lekshmi, Nandi and Antia (2018) concerning their study of correlation between hemispheric asymmetry in solar photospheric torsional oscillations (i.e., residuals of solar differential rotation rate) and the photospheric magnetic flux as well as the sunspot numbers. They found a very strong and statistically significant (about 99.9 %) cross-correlation between asymmetry in these torsional oscillations and asymmetry in two solar activity indicators (solar surface magnetic flux and sunspot numbers). They concluded that asymmetry in near-surface torsional oscillation velocity leads the asymmetries corresponding to magnetic flux and sunspot numbers with average time of 1.41 and 1.18 years, respectively, covering the declining phase of solar cycle 23 and complete solar cycle 24. However, our results illustrates that during the period from 2008 to 2018 covering the solar cycle 24, hemispheric asymmetry in solar rotation rate leads sunspot numbers with an average shift of about 1.56 years. Hence, we conjecture from our analysis that the formation of sunspots (or the solar activity) could be triggered by the hemispheric asymmetry in solar rotation rate. Here, it appears that asymmetry could be promoting the twisting of magnetic field lines which creates the sunspots in the solar photosphere. Additionally, the differential rotation as well as asymmetry in differential rotation of the solar interior and photosphere must be related to those regions that we are investigating using EUV observations. Outer parts of Sun's atmosphere namely transition region and corona rotate because there is rotation in the solar photosphere and interior of the Sun. Thus differential rotation and asymmetry seen here; indicate a possibility that in solar interior and photosphere rotational differentiability and hemispheric asymmetry may be similar or more. Our results (a very high cross-correlation of AI with SSNs and EUV flux) suggest that differential rotation and hemispheric asymmetry seem to play a key role in the formation of sunspot number and solar activity. We have no doubt that our findings provide clues to fellow specialists who work towards better forecast and understand solar variability.

## ACKNOWLEDGEMENTS

The authors acknowledge the use of data (30.4, 19.5 and 28.4 nm) from the *Solar Terrestrial Relations Observatory (STEREO)* for the period 2008–2018. These were acquired from the webpage of the *STEREO* which is a NASA's operated mission. We also acknowledge the webpage of SILSO-SIDC from which annual sunspots numbers from 2008 to 2018 has been used in this analysis. The research at Udaipur Solar Observatory (USO), Physical Research Laboratory; Udaipur is supported by Department of Space, Government of India. We also acknowledge the various supports for this research work provided by Department of Physics, Chaudhary Charan Singh University, Meerut, India. We are also thankful to the referee for useful comments and suggestions to improve this manuscript.

## DATA AVAILABILITY

The data underlying this article will be shared on reasonable request to the corresponding author.

## REFERENCES

Brun A.S., (2004), **Sol. Phys,** 220, 333.
Brun A.S., Miesch M.S., Toomre J., (2004), **APJ,** 614, 1073.
Cameron R.H., Dikpati M., Brandenburg A., (2017), **Space Sci. Rev,** 210, 367.
Chandra S., Vats H. O., Iyer K. N., (2010), **MNRAS,** 407, 1108.
Chandra S., Vats H.O.,(2011), **MNRAS,** 414, 3158.
Georgieva K., Kirov B., Javaraiah J., Krasteva R., (2005), **Planet. Space Sci, 53, 197.**
Hansen R., & Hansen S. (1975), **Sol. Phys,** 44, 225.
Hathaway D. H., Wilson R. M., (2004), **Sol. Phys,** 224, 5.
Hathaway D. H., Wilson R. M., Reichmann E. J., (1999), **J. Geophys. Res,** 104, 22375.
Hiremath K. M., Mandi P. I., (2004), **New Astron,** 9, 651.
Joshi B., & Joshi A., (2004), **Sol. Phys,** 219, 343.
Jurdana- Sepic R., Brajsa R., Wohl H., Hanslmeier A., Poljancic I., Svalgaard L., Gissot S.F., (2011), **A&A,** 534, A17.
Lanza A.F., (2007), **A&A,** 471, 1011.
Lekshmi B., Nandy D., Antia H.M., (2018), APJ, 861,121.
Li K.J., Feng W., Shi X.J., Xie J.L., Gao P.X., Liang H.F., (2014), **Sol. Phys,** 289, 759.
Rozelot J. P., (2001), **J. Atmos. Phys,** 63, 375.
Sharma J., Kumar B., Malik A.K., Vats H.O., (2020), **MNRAS,** 492, 5391.
Sokoloff D., Nesme-Ribes E., (1994), **A&A,** 288, 293.
Vats H. O., Cecatto J. R., Mehta M., Sawant H. S., Neri J. A. C. F., (2001), **ApJ,** 548, 87
Vats H. O., Chandra S., (2011), **MNRAS,** 000, 1-5.
Verma V. K. (1993), **ApJ,** 403, 797.

Verma V. K., (2000), **Sol. Phys,** 194, 87.

Wuelser J., Lemen J., Tarbell T., Wolfson C. (2004), **Telescopes and Instrumentation for Solar Astrophysics,** 5171.